\documentclass{webofc}
\usepackage[varg]{txfonts}   
\begin{document}
\title{Collective Dynamics - theoretical overview}
%
%

\author{
\firstname{Yuuka} \lastname{Kanakubo}\inst{1,2}\fnsep\thanks{\email{yuka.y.kanakubo@jyu.fi}}
}

\institute{
University of Jyväskylä, Department of Physics, P.O. Box 35, 40014 University of Jyväskylä, Finland
\and 
Helsinki Institute of Physics, P.O. Box 64, 00014 University of Helsinki, Finland
}

\abstract{%
I overview the recent progress of phenomenological studies exploring collective dynamics in relativistic nuclear collisions
to understand various QCD properties.
Originally, collectivity was interpreted as a manifestation of the hydrodynamic behaviour of the QGP as a response to the initial collision geometry.
Over the past decade, however, particularly following the experimental observation of collectivity in small colliding systems, 
pioneering studies have demonstrated the possibility of other interpretations.
In this talk, I highlight recent studies aimed at understanding various QCD properties at different collision stages through the lens of collectivity and emphasize the importance of establishing Monte Carlo event generators for relativistic nuclear collisions.
}
\maketitle

\vspace{-20pt}
\section{Introduction}
\label{intro}
Collectivity, a momentum azimuthal anisotropy of final state hadrons, has been interpreted as a hydrodynamic behaviour of the QGP in relativistic heavy-ion (A--A) collisions \cite{Ollitrault:1992bk}.
Geometrical anisotropy in the initial profile of quark-gluon plasma (QGP) gives an anisotropic pressure gradient, which generates flow velocity as a consequence of hydrodynamic response \cite{Molnar:2001ux,Kolb:2003dz}.
The hydrodynamic response is controlled by the viscous coefficients of QGP and the equation of state.
Hence, the comparison of momentum anisotropy between experimental results and hydrodynamic model calculations is anticipated to provide insights into the properties of QGP.
This assumes that collectivity arises solely from the hydrodynamic expansion against the initial geometrical anisotropy of QGP originating from the collision geometry.
However, the above interpretation of the emergence of collectivity has undergone a significant change particularly 
following the experimental observation of collectivity in proton--proton (p--p) collisions or proton--ion (p--A) collisions.
These collisions were previously regarded as systems too small to produce QGP \cite{DobrigkeitChinellato:2023qm} and do not have such a clear collision geometry compared to A--A collisions.

\vspace{-5pt}
\section{Origins of collectivity}
\label{origin}
A naive question to pose here is
whether the mechanism giving rise to collectivity in small systems is the same as in heavy-ion collisions or not.
This opened up an opportunity to excavate other possible origins of collectivity in small systems \cite{Dusling:2015gta,Nagle:2018nvi,Schenke:2021mxx}.
For instance, hot spots of a gluonic profile of a proton as a seed of geometrical anisotropy in the initial state \cite{Schenke:2021mxx}, collectivity arising from pre-equilibrium evolution \cite{Bozek:2022cjj}, or
collectivity from string interactions in the initial state and hadronic rescattering in the final state \cite{Bierlich:2021poz} etc.
Also, it should be noted that not only hydrodynamic models \cite{Schenke:2020mbo,Zhao:2020wcd,Zhao:2022ugy} but also 
transport models \cite{Zhao:2021bef,Oliva:2022rsv} reasonably describe the collectivity observed in experimental data.
Given that various physical mechanisms can potentially explain the observed collectivity in experiments, qualitative and quantitative discrimination of contributions from those physical mechanisms is essential for a comprehensive understanding of dynamics in each collision stage.

An illustrative example of the discussion on the origin of collectivity can be seen in the recent ATLAS experimental data in ultra-peripheral collisions (UPC) \cite{ATLAS:2021jhn}.
The finite azimuthal anisotropy coefficient, $v_n$, was reported as a function of transverse momentum $p_T$ in high-multiplicity UPC events.
There are two theoretical calculations which try to explain these experimental data, a calculation with
color glass condensate (CGC) \cite{Shi:2020djm} and hydrodynamics \cite{Zhao:2022ayk}.
Both show a reasonable agreement of $v_2$ with the data.
In the former/latter calculation, the collectivity arises as a consequence of the initial/final state effect \cite{Bozek:2016yoj,Giacalone:2020byk,Lim:2021auv}.
Whether the collectivity in small systems comes from the initial or final state effect is a long-standing problem and still an open question.

The recent study with the (3+1)D IP-Glasma framework \cite{Schenke:2022mjv} tries to 
tackle this question.
In this framework, the longitudinal structure of incoming nuclear gluon distributions is obtained by the JIMWLK equation. 
The results demonstrate that the initial momentum anisotropy has a short-range correlation in rapidity while the geometrical anisotropy has a long-range correlation.
This implies that the geometrical anisotropy can be a dominant contribution to the collectivity 
considering that the collectivity in small-system experimental data has a long-range correlation.

\vspace{-5pt}
\section{Nuclear structure}
\label{NuclearStructure}
The nuclear structure holds information on the spatial correlation of nucleons inside of a nucleus.
Recent studies show that the nuclear structure can be accessible also through the flow observables in relativistic heavy-ion collisions \cite{Giacalone:2023cet}.
For instance, the difference of $v_2$ and $v_3$ in Ru$^{\mathrm{96}}$+Ru$^{\mathrm{96}}$ and Zr$^{\mathrm{96}}$+Zr$^{\mathrm{96}}$ collisions can be explained by the difference of quadrupole $\beta_2$ and octupole $\beta_3$ deformation in Ru$^{\mathrm{96}}$ and Zr$^{\mathrm{96}}$ \cite{Jia:2021tzt,Zhang:2021kxj}.
The deformation parameters such as $\beta_2$ and $\beta_3$ are encoded in the generalized Woods-Saxon distributions.
The information of the deformation is reflected in the initial geometry of QGP, which affects $v_2$ and $v_3$ in the final state.

A recent study emphasizes the potential necessity of using Woods-Saxon parameters directly obtained from a rigorous calculation based on many-body quantum mechanics for collision systems with deformed nuclei \cite{Ryssens:2023fkv}.
One can extract the deformation parameters by fitting the generalized Woods-Saxon distribution to the three-dimensional density profile from Skyrme density functional calculations.
The calculation with hydrodynamics using the best-fit parameters shows a better agreement with experimental data for the $v_2\{2\}$ ratio between Au+Au and U+U collisions compared to the result with the previously obtained parameters.

There are still a lot of remaining tasks to be addressed in this study: the (in)consistency of the nuclear structure between different energy scales \cite{Mantysaari:2023prg,Mantysaari:2023qsq,Singh:2023qm}, computational cost of initial profiles with deformed nuclei \cite{Luzum:2023gwy}, or non-equilibrium effects in the flow analysis \cite{Kanakubo:2022ual} etc.
However, it can be clearly seen that the studies on nuclear structure have paved the way to investigate a connection between high and low-energy nuclear physics.

\vspace{-5pt}
\section{Developments on statistical analysis}
\label{statistical}
One of the primary goals of constructing dynamical models is to extract physical parameters through comparisons between the model and data. Recently, Bayesian parameter estimation has become the mainstream approach for this purpose, propelled by pioneering works in the field \cite{Pratt:2015zsa,Bernhard:2016tnd}.
Although originally the transport coefficients of 
QGP were the main target to study in the Bayesian analysis, 
the possibility of investigating other physical phenomena has been explored too.
For instance, one can study the effect of neutron skins \cite{Giacalone:2023cet} or deuteron production mechanisms \cite{JETSCAPE:2022cob}.
Also, the Bayesian parameter estimation can be extended to a viscous anisotropic hydrodynamic framework \cite{Liyanage:2023nds,Heinz:2023qm}, where 
$\eta/s(T)$ and $\zeta/s(T)$ can be constrained up to higher temperatures compared to a framework with standard viscous hydrodynamics.

However, despite the growing popularity of Bayesian parameter estimation, it is crucial to acknowledge that there are always inherent uncertainties in a theoretical model. 
Accurately estimating these uncertainties can be challenging.
In the previous results with Bayesian analysis, there 
has been a common issue: the extracted nucleon width was $w\sim 0.8-1.0$ fm,
which was quite large compared to, for example, the proton radius that is generally expected.
This issue is reconciled by including total hadronic nucleus-nucleus cross-section $\sigma_{AA}$
in the Bayesian analysis \cite{Nijs:2022rme} utilizing the fact that $\sigma_{AA}$ is sensitive to $w$.
The results show that the estimated $w$ becomes below 0.7 fm and the $v_2\{2\}^2$-$\langle p_T \rangle$ correlation with the best-fit parameters shows significant improvement in the description of experimental data.

The significant computational cost of dynamical models is another problem to be addressed
to achieve precise model-to-data comparisons.
A potential solution to this issue is the application of deep convolutional neural networks, 
which provides robust assistance for the model prediction \cite{Hirvonen:2023lqy}.
The neural network is trained to predict the hydrodynamic results event by event, $e.g.$ flow coefficients,  average transverse momentum and charged particle multiplicities, from an initial energy density profile.
It is shown that the trained neural network gives reliable predictions of hydrodynamic results with just a fraction of computational time.
Sufficient statistics are easily attained compared to the calculation with full evolution of hydrodynamics, which significantly reduces the statistical errors of the computed flow observables, and of the rarest flow correlators in particular.

\vspace{-5pt}
\section{Monte Carlo event generators for relativistic nuclear collisions}
\label{MCgenerator}
As mentioned above, model-to-data comparisons in relativistic nuclear collisions 
have become sophisticated because of the development of powerful statistical methods.
Nevertheless, a notable question remains: ``Are the results from dynamical frameworks comparable enough for direct comparisons with experimental data?''
There are two missing points in conventional hybrid models.
First, the description is limited to particle production in low $p_T$ regions 
while it is hard to disentangle the soft and hard production when there is an interplay between them \cite{Pablos:2022piv}.
Secondly, there is a lack of serious consideration for energy-momentum conservation on an event-by-event basis, particularly in obtaining the initial state and during the conversion of fluids to particles at the end of the hydrodynamic simulation.
Both have been considered essential features for general-purpose Monte-Carlo event generators in high-energy physics \cite{Buckley:2011ms}.

\begin{figure}[h]
    \centering
    \includegraphics[bb= 0 0 634 446, width=0.48\textwidth]{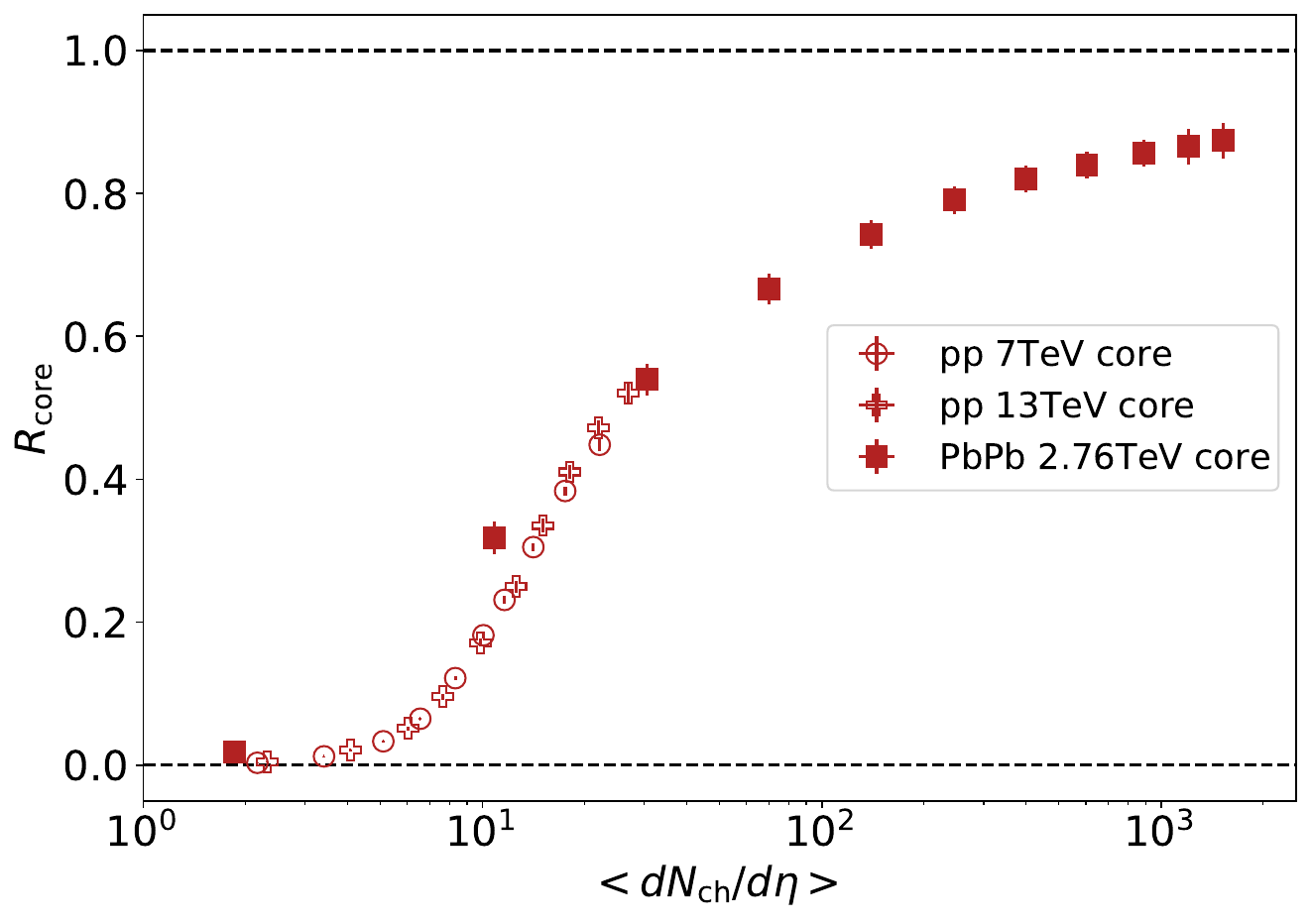}
    \label{fig:result}
    \includegraphics[bb= 0 0 634 446, width=0.48\textwidth]{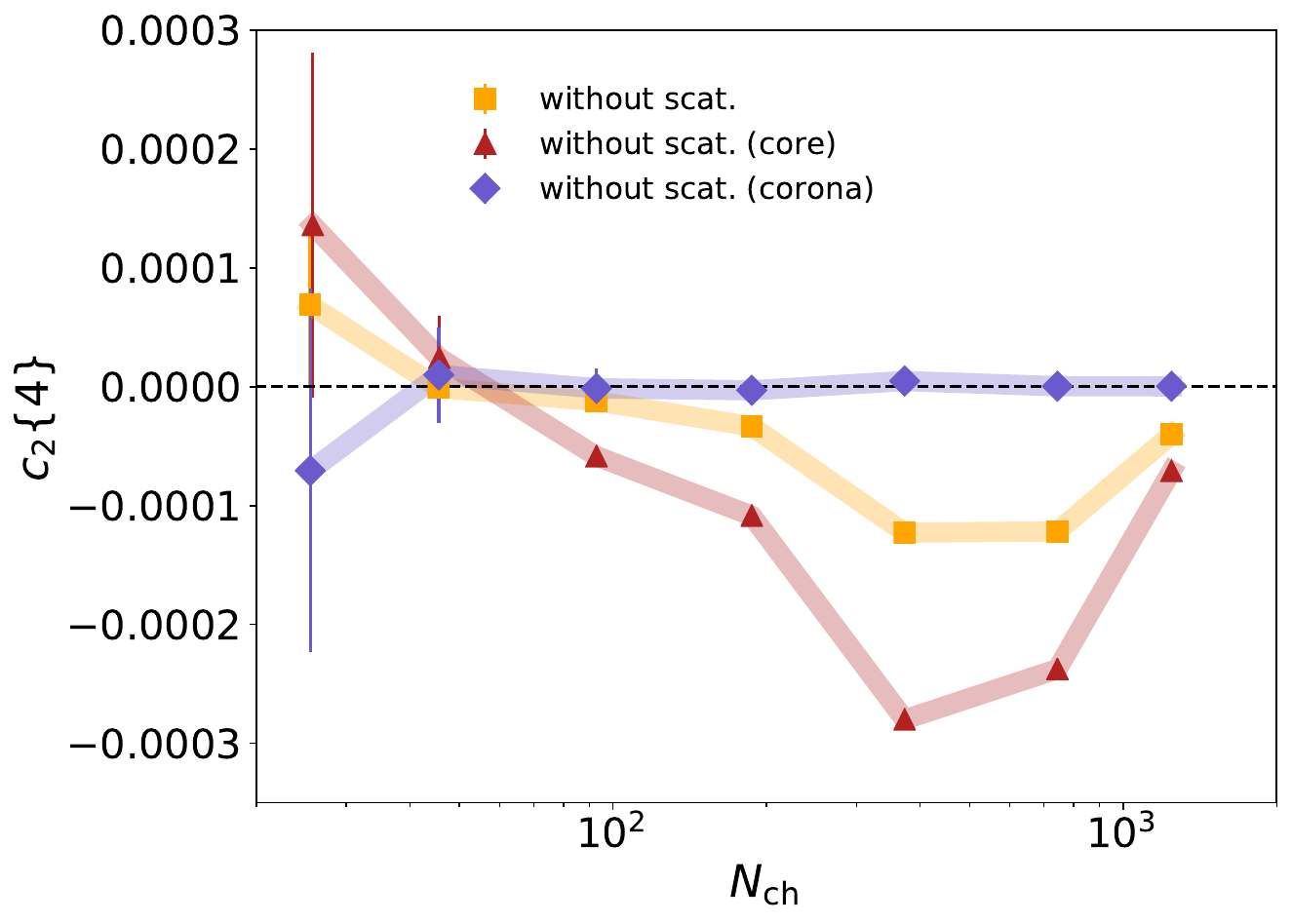}
    \caption{(Left) Fraction of hadrons produced from the QGP fluids (core) at midrapidity as a function of charged-particle multiplicity 
    extracted from the dynamical core--corona initialization framework \cite{Kanakubo:2021qcw}.
    Results for p--p collisions at $\sqrt{s} = 7$, $13$ TeV, and Pb--Pb collisions at $\sqrt{s_{NN}}=2.76$ TeV are shown.
    (Right) Four-particle cumulants $c_2\{4\}$ for charged hadrons as a function of the number of charged hadrons $N_{\mathrm{ch}}$ in Pb--Pb collisions at $\sqrt{s_{NN}}=2.76$ TeV \cite{Kanakubo:2022ual}. Results calculated only with hadronic production from the core component (triangles), the corona components (diamonds), and the inclusive hadrons (squares) are shown.
    }
    \label{fig:result2}
\vspace{-20pt}
\end{figure}

The latest version of EPOS, EPOS4 \cite{Werner:2023zvo,Werner:2023fne,Werner:2023jps},
tries to tackle this problem.
The key feature in the EPOS is the core--corona picture -- a two-component picture of equilibrated (core) and non-equilibrated (corona) matter.
By consistently including the particle production from corona components in hydrodynamics, 
the applicable range of the model is extended to high $p_T$ regions and small systems
where local equilibrium is not likely attained.
Notably, the initial state model with parallel scattering formalism has been significantly updated in the EPOS4.
Factorization/binary scaling is attained at high $p_T$ regions 
and saturation regulates particle production at low $p_T$ regions,
which makes the model able to describe soft and hard production within a single framework under the restriction of energy-momentum conservation.
Also, it should be noted that energy-momentum and charge conservation are achieved event-by-event in the conversion of fluids to particles by implementing a microcanonical sampling.
Taking into account all the aforementioned features,
the EPOS4 is an MC-event generator modelled to cover the physics from lepton--lepton to heavy-ion collisions with possible QGP formation.

The dynamical core--corona initialization framework (DCCI2) \cite{Kanakubo:2021qcw,Kanakubo:2022ual} is established by implementing the concept of the core--corona picture in the dynamical initialization framework.
In the DCCI2, the source term of hydrodynamics is defined so that energy-momentum deposition of non-equilibrated partons is proportional to their scattering rate evaluated from the mean free path.
Because of the core--corona picture, the framework describes the particle yield ratios as a function of multiplicity from small systems to heavy-ion collisions similarly to the EPOS4.
Figure \ref{fig:result} (left) shows the fraction of the number of hadrons produced from the core component, $R_{\mathrm{core}}$, in the DCCI2.
The clear scaling of $R_{\mathrm{core}}$ with multiplicity is seen for p--p collisions at $\sqrt{s} = 7$, $13$ TeV, and Pb--Pb collisions at $\sqrt{s_{NN}}=2.76$ TeV.
The onset of QGP dominance appears to be at $\langle dN_{\mathrm{ch}}/d\eta_s \rangle \sim 10$-$20$, which is roughly consistent with the result from EPOS4 \cite{Werner:2023zvo}.
It is also important to notice that the contribution from the corona components is not negligible in heavy-ion collisions.

In Fig.~\ref{fig:result2} (right), the effect of the corona component on $c_2\{4\}$ is investigated in Pb--Pb collisions. Comparisons of $c_2\{4\}$ calculated only with hadronic production from the core component and with the inclusive hadronic production (core+corona) reveal the non-equilibrium correction on $c_2\{4\}$ that is purely obtained from hydrodynamics. 
The correction is non-negligible especially in mid-central collisions even if $c_2\{4\}\sim 0$ in hadronic production from the corona component. 
This means that one cannot simply compare the absolute value of $c_2\{4\}$ obtained from pure hydrodynamic models with experimental data due to the existence of non-equilibrium correction
even if the contribution of ``non-flow'' correlation is subtracted in $c_2\{4\}$.
Therefore, the result urges that the quantitative model-to-data comparisons of flow observable
should be performed considering both equilibrated and non-equilibrated components, ideally, within a MC event generator.

There are pioneering works towards the establishment of MC-event generators for QGP studies,
$e.g.$ (3+1)D hydrodynamics with MC-EKRT minijets \cite{MCEKRT},
sampling of thermal hadrons with energy-momentum and charge conservation \cite{Oliinychenko:2019zfk,Werner:2023jps},
the frameworks with core--corona pictures as discussed above \cite{Werner:2023zvo,Kanakubo:2021qcw} etc.
Ultimately, a full 3D Bayesian parameter estimation with MC-event generators  \cite{Auvinen:2017fjw,Shen:2023qm,Mankolli:2023qm} would make it possible to
perform a quantitative extraction of physical parameters from rigorous model-to-data comparisons.
It also should be noted that the extension of the 
applicability range of the models should be extended to lower collision energies \cite{Du:2022yok,Cimerman:2023hjw,Pihan:2023qm,Shen:2023qm,Almaalol:2022pjc} 
for the investigation of the structure of the full QCD phase diagram.

\vspace{-5pt}
\section{Conclusions}
\label{conclusions}
In relativistic nuclear collisions, collectivity can arise from various physical mechanisms. This implies a substantial opportunity to explore diverse aspects of QCD physics through the study of collectivity.
On the other hand, this also indicates that one should disentangle the contribution from different physics and needs to quantitatively identify each contribution in collectivity.
Developing Monte Carlo event generators for relativistic nuclear collisions would be crucial 
for such quantitative model-to-data comparisons with a full 3D Bayesian analysis.

\vspace{10pt}
{\bf{Acknowledgement}}
I would like to thank Giuliano Giacalone, Henry Hirvonen, Mikko Kuha, Heikki Mäntysaari, Govert Njis, Wouter Ryssens, Wilke van der Schee, Pragya Singh, and all members of the Centre of Excellence in Quark Matter in Jyväskylä, for discussions.
Our research was funded as a part of the Center of Excellence in Quark Matter of the Academy of Finland (Project No. 346325), the European Research Council Project No. ERC-2018-ADG835105 YoctoLHC, the Academy of Finland Project No. 330448, and the European Union’s Horizon 2020 research and innovation program under grant agreement No. 824093 (STRONG-2020).

\vspace{-10pt}
%
\bibliography{bibs}

\end{document}